\documentclass[english]{emulateapj}
\usepackage{color}

\makeatletter
\renewcommand\@biblabel[1]{#1}

\@ifundefined{textcolor}{}
{%
 \definecolor{BLACK}{gray}{0}
 \definecolor{WHITE}{gray}{1}
 \definecolor{RED}{rgb}{1,0,0}
 \definecolor{GREEN}{rgb}{0,1,0}
 \definecolor{BLUE}{rgb}{0,0,1}
 \definecolor{CYAN}{cmyk}{1,0,0,0}
 \definecolor{MAGENTA}{cmyk}{0,1,0,0}
 \definecolor{YELLOW}{cmyk}{0,0,1,0}
 }

\makeatother

\usepackage[normalem]{ulem}  

\newcommand{\source}{IGR~J17062$-$6143}

\usepackage{babel}

\shorttitle{\source{} is a millisecond pulsar}
\shortauthors{Strohmayer \& Keek}

\begin{document}

\title{IGR J17062$-$6143 is an Accreting Millisecond X-ray Pulsar}

\author{Tod Strohmayer$^1$ and Laurens Keek$^2$ \\ {\normalfont
    $^1$Astrophysics Science Division and Joint Space-Science
    Institute, NASA Goddard Space Flight Center, Greenbelt, MD
    20771, USA} \\ {\normalfont $^2$X-ray Astrophysics Laboratory,
    NASA/GSFC \& CRESST and the Department of Astronomy, University of
    Maryland, College Park, MD 20742, USA}}

\begin{abstract}

We present the discovery of 163.65 Hz X-ray pulsations from IGR
J17062$-$6143 in the only observation obtained from the source with
the {\it Rossi X-ray Timing Explorer}.  This detection makes IGR
J17062$-$6143 the lowest-frequency accreting millisecond X-ray pulsar
presently known. The pulsations are detected in the 2 - 12 keV band
with an overall significance of $4.3 \sigma$, and an observed pulsed
amplitude of $5.54 \pm 0.67\%$ (in this band).  Both dynamic power
spectral and coherent phase timing analysis indicate that the
pulsation frequency is decreasing during the $\approx1.2$ ks
observation in a manner consistent with orbital motion of the neutron
star. Because the observation interval is short, we cannot precisely
measure the orbital period; however, periods shorter than 17 minutes
are excluded at $90\%$ confidence. For the range of acceptable
circular orbits the inferred binary mass function substantially
overlaps the observed range for the AMXP population as a whole.


\end{abstract} 

\section{Introduction}

\label{sec:introduction} There is now compelling evidence that millisecond 
pulsars are ``recycled,'' meaning that they have acquired their rapid
spins via accretion during their binary evolution. A compelling
demonstration of the recycling scenario was provided by the discovery
in 1998 of the first accreting millisecond X-ray pulsar (AMXP), SAX
J1808.4$-$3658 (hereafter J1808). This system comprises a 401 Hz
pulsar in a 2.01 hr binary (Wijnands \& van der Klis 1998; Chakrabarty
\& Morgan 1998). It remains the ``prototype,'' and best-studied member
of the class, and exhibits all of the defining properties of AMXPs:
short orbital periods (less than a few hours, and often less than an
hour), low time-averaged accretion rates, and very low mass donors
($\leqslant 0.2 M_{\odot}$).  It has now been eighteen years since the
discovery of J1808, and still only 16 such sources have been
identified (see Patruno \& Watts 2012 for a review), the most recent
being {\it Monitor of All-sky X-ray Image} (MAXI) J0911$-$655 (Sanna
et al. 2016).

In this Letter, we present evidence based on an observation performed
in 2008 May with the {\it Rossi X-ray Timing Explorer} (RXTE; Bradt et
al. 1993) that \source{} is an AMXP with a pulsation frequency of
163.65 Hz.  The source (also classified as SWIFT~J1706.6$-$6146) is an
accreting neutron star binary, first observed during an outburst in
2006 (Churazov et al. 2007; Ricci et al. 2008; Remillard \& Levine
2008). The outburst has persisted at a low flux level since then, and
the source has also produced several bright X-ray bursts, confirming
that the accretor is a neutron star. The first of these was observed
by {\it Swift} in 2012 following an on-board trigger from the {\it
  Swift/BAT} (Degenaar et al. 2013), and more recently, a
long-duration burst, possibly powered by a thick helium layer (Keek et
al. 2016), was detected by the MAXI all-sky monitor (Negoro et
al. 2015). Both events exhibited signs that the bursts had a strong
impact on the accretion environment around the neutron star. This
included variability in the X-ray light curve as well as reflection
features in the spectrum.  Reflection signatures were also detected by
{\it NuSTAR} in observations of the persistent flux (Degenaar et
al. 2017; Keek et al. 2016). They indicate that the accretion disk is
truncated at $\sim 10^2\,R_\mathrm{g}$, where $R_\mathrm{g}=GM/c^2$ is
the gravitational radius. A relatively strong magnetic field could
potentially provide the support for the truncated inner disk.

We first present the results of our pulsation search of \source{} and
describe the discovery of pulsations (Section~\ref{sec:pulsation}). We
show that the pulse frequency varies with time in a manner consistent
with orbital motion of the neutron star. We further use coherent
timing methods to place a lower limit on the orbital period of about
17 minutes. The energy spectrum is analyzed to quantify the mass accretion
rate at the time of the observation. Next, we estimate the magnetic
field strength, and we discuss whether it supports the picture of the
accretion geometry that was inferred by the reflection spectra
(Section~\ref{sec:discussion}). 


\section{Pulsation Search with RXTE}

\label{sec:pulsation} Perhaps surprisingly, {\it RXTE} observed 
\source{} only once, in 2008 May for a total of $\approx 1200$ s
(obsid: 93437-01-01-00). We used this single observation to search for
pulsations.  In addition to the standard Proportional Counter Array
(PCA) data modes, high time resolution data were acquired in the {\it
  E\_125us\_64M\_0\_1s} event mode, which provides a time resolution
of 1/8192 s in 64 energy bins.  For this observation only two of the
five Proportional Counter Units (PCUs) comprising the full PCA array
were active (PCUs 0 and 2, in the 0-4 numbering scheme).  We used the
standard {\it RXTE} data analysis tools to extract light curves,
spectra, and an estimate of the background spectrum during the
observation.  We used the tool {\it faxbary} to compute barycentric
arrival times for all events.  For this purpose we used the source
position, $R.A. = 256.5677$, decl. $ = -61.7113$, from Ricci et
al. (2008).  We used the barycentered events to construct light curves
with 1/8192 second resolution (for a Nyquist frequency of 4096 Hz) and
the observation had a useful exposure of 1168 s.

Since the background provides a significant fraction of the total
count rate, we carried out a pulsation search in two energy bands.
The first used all counts across the entire PCA response.  For the
second we found the energy channel below which the source count rate
is greater than the background counting rate and only used events with
energy less than or equal to this value. We found that channel 16 (of
the 64 bins of the data mode) satisfied this condition, so we produced
a light curve using only events in channels 16 and lower.  For this
data mode and observation epoch, that channel corresponds to photons
with energies less than or equal to $\approx 12$ keV.  Figure 1 shows
the resulting light curve at 1 s resolution in the $2 - 12$ keV band,
as well as the estimated background (red curve) at 16 s resolution
computed from {\it pcabackest}.

We computed Leahy-normalized power spectra for each light curve and
searched for pulsations in the frequency range from 10 to 2048 Hz.
Since orbital motion can induce frequency drifts, we carried out a
``hierarchical'' search by averaging adjacent Fourier powers in
frequency space by a factor, $W$, and also searched the averaged
spectra. We used values of $W = 1, 2, 4, 8, 16$, for a total of 10
power spectra searched. We found a strong pulsation candidate at
163.655 Hz in the power spectrum computed using only the 2 - 12 keV
events and averaged by a factor of $W=8$. Figure 2 shows this power
spectrum, revealing the strong peak near 163.65 Hz. The putative
signal peak has a power value of 11.646.

In such a ``hierarchical'' search, the rigorous way to assess
significances is via Monte Carlo simulations because the distribution
of noise powers is different for each averaged power spectrum (van der
Klis 1989). First, we note that in the absence of signal power (the
null hypothesis) the power values in an averaged power spectrum are
distributed such that the probability that a power $P_j$ will exceed
some value $P$ is given by $Q_{\chi^2} (WP, 2W)$, where $Q_{\chi^2}$
is the integral probability of the $\chi^2$ distribution with $2W$
degrees of freedom. As above, $W$ is the averaging factor, and with
$W=1$ the familiar $\chi^2$ distribution with 2 degrees of freedom is
obtained.  Thus, the peak power value of 11.646 has a chance
occurrence, per trial, given by the probability of obtaining a value
of $8 \times 11.646 = 93.17$ from a $\chi^2$ distribution with 16
degrees of freedom. This single trial probability is $6.5 \times
10^{-13}$. To arrive at our significance estimate, we carried out Monte
Carlo simulations that precisely mimic our search procedures.  We
simulate two light curves with the same mean count rates as our
observed light curves. To do this, we generate Poisson realizations
using the same time bin size and light curve duration as observed.  We
then compute averaged power spectra for each simulated light curve
using $W = 1, 2, 4, 8, 16$, and we identify in each spectrum the
frequency bins within our $10 - 2048$ Hz search range.  For each of
these bins we determine the single trial probability using the
appropriate $Q_{\chi^2} (WP, 2W)$ distribution, and we test whether
any bin has a single trial probability less than or equal to the
observed value ($6.5 \times 10^{-13}$) of the putative signal peak.
We repeat the process many times to determine how often a single trial
probability this small is achieved in at least one of the light
curves. We also determined the number of times that several larger
(less significant) single trial probabilities were exceeded, and in
this way we determined the $3\sigma$ confidence detection threshold,
which is plotted as the horizontal dashed line in Figure 2. From
$5\times 10^5$ simulations we find a significance of $\approx
1.6\times 10^{-5}$ for the 163.65 Hz pulsation, which is better than a
$4\sigma$ detection.


As a consistency check, we used the $W=8$ power spectrum (see Figure
2) in the range from 2048 to 4096 Hz (where no signal power is
expected, and we did not search) to explore the distribution of powers
and see whether it tracks the expected distribution. We found that the
distribution of noise powers in this power spectrum matches well with
the expected $Q_{\chi^2} (8P , 16)$ distribution, giving us additional
confidence that our significance estimate is robust.

\subsection{Evidence for Orbital Motion}

We next computed dynamic power spectra in order to determine if any
secular variations in the pulsation frequency, as would be produced by
orbital motion of the neutron star, could be discerned.  Since the
signal strength is relatively modest, and there is only a single {\it
  RXTE} orbit of data, we used a relatively long time interval of 512
s to compute power spectra, and then shifted the center of the
interval by 16 s to compute a dynamic spectrum.  Figure 3 shows a map
of contours of constant Fourier power versus time and frequency from
this analysis. The contours are drawn at the time corresponding to the
center of each time interval used to compute a power spectrum. Nine
contours are drawn at power levels of 20, 22, 24, 26, 28, 30, 32, 36,
and 40.  Because the segments overlap, each power spectral measurement
is not completely independent; nevertheless, the time evolution of the
Fourier power is indicative of a decrease in pulsation frequency
during the observation, which would not be unexpected for orbital
motion of the neutron star.  The apparent variation in frequency of
$\Delta\nu \approx 0.002$ Hz is likely a lower limit to the full range
of any orbitally induced variations.  This corresponds to a lower
limit on the projected neutron star velocity of $v/c = \Delta\nu /
\nu_0 \approx 1.22\times 10^{-5}$ or $\approx 3.7$ km s$^{-1}$, which
is consistent with measurements of other AMXPs (see, for example,
Markwardt et al. 2002).

To further explore this, we also carried out a coherent timing
analysis (see Buccheri et al. 1983; Strohmayer \& Markwardt
2002). Since all other AMXPs are adequately described by circular
orbits, we used a circular orbit model to describe the time evolution
of the average pulsation phase through the observation interval.  This
model provides a statistically acceptable description of the phase
evolution.  Further, we also show using this method that a constant
frequency model is not consistent with the phase evolution. Figure 4
compares phase timing residuals for these two models.  The black
square symbols show the phase residuals from the best constant
frequency model, which gives an unacceptably high $\chi^2$ value of
$40.1$ for 5 degrees of freedom.  Moreover, the overall trend in these
residuals of a concave-up, parabolic shape is strongly indicative of a
spin-down of the pulsation frequency, consistent with the dynamic
power spectrum (Figure 3). Additionally, in Figure 4, the red diamond
symbols show the phase residuals for a circular orbit with a period of
30 minutes. This model gives a good fit to the phase data, with a
minimum $\chi^2 = 1.1$ for 2 degrees of freedom.  While this orbital
model provides an acceptable fit to the phase evolution, because the
observation is so short, we cannot precisely determine the orbital
period. Indeed, we find that orbits with periods longward of about 17
minutes can all produce acceptable phase residuals; however, we find
that the minimum $\chi^2$ values rise to unacceptable levels as the
orbital period is further reduced, and we find a 90\% confidence
($\Delta \chi^2 = 2.71$) lower limit of 17 minutes.  We used the
best-fitting orbit model with a 30 minute period to phase-fold all the
2 - 12 keV events in order to study the pulse profile and amplitude.
The resulting profile is well described by the model, $A + B\sin (\phi
- \phi_0)$. Based on this model, and after subtraction of the
background, we find an average source pulsed amplitude of $B/A = 9.4
\pm 1.1 \%$. We also subdivided the observation interval into quarters
and measured the amplitude in each, but found no evidence for
significant variations in pulsed amplitude.

For the range of acceptable circular orbits we can determine the mass
function, $f_x = ( (a\sin i)^3 \omega_{orb}^2 )/G$, where $a\sin i$
and $\omega_{orb}$ are the projected neutron star orbital radius and
frequency, respectively.  We find that the mass function increases
monotonically from $2 \times 10^{-8}$ at a period of 30 minutes to
$2.5 \times 10^{-6}$ for a period of 100 minutes. This range
substantially overlaps with the observed range for other AMXPs (see
Patruno \& Watts 2012), but, interestingly, if the period is close to
30 minutes, then the mass function would likely be the smallest yet
observed and would correspond to a minimum companion mass less than
$0.004 M_{\odot}$, the smallest to date.  Thus, additional
observations to pin down the orbital period would be quite important.

\subsection{Analysis of the Spectral Energy Distribution}

We extracted a single PCA energy spectrum from the full observation
interval in the $3-20$ ~keV energy range using Standard 2 data from
the top layer of PCUs $0$ and $2$. The spectrum was analyzed using
{\sc XSPEC} version 12.9.0n (Arnaud 1996). It is well fit by an
absorbed power-law model (Figure~\ref{fig:spectrum}). We employ the
T\"ubingen$-$Boulder model with abundances from Wilms et al. (2000) to
model the effects of interstellar absorption. The absorption column is
fixed to $N_{\mathrm{H}}=1.58\times10^{21}\,\mathrm{cm^{-2}}$ (Keek et
al. 2016).  We find a power-law photon index of $\Gamma=2.21\pm 0.03$
and an unabsorbed $3-20$~keV flux of $F=(1.28\pm 0.02)\times
10^{-10}\mathrm{erg\,s^{-1}\,cm^{-2}}$.  $\Gamma$ is only
$1.7\,\sigma$ larger than the value measured with {\it Chandra} and
{\it NuSTAR} in 2014 and 2015, respectively (Keek et al. 2016; see
also Degenaar et al. 2017), and $F$ is $2.5$ times larger than for the
{\it NuSTAR} spectrum in the same band. A previous measurement of the
source flux during an Eddington-limited, photospheric radius expansion
(PRE) episode in a thermonuclear burst suggests a mass accretion rate
of $0.6\%$ of the Eddington limit at the time of the PCA observation
(Keek et al. 2016).  We note that for an assumed Eddington luminosity
of $2 \times 10^{38}$ erg s$^{-1}$, the implied source distance is
$\approx 8.8$ kpc. However, the bolometric correction may be different
for the PCA spectrum because the blackbody component that was present
in 2014 and 2015 is not visible in the PCA spectrum. Perhaps the
blackbody has a smaller normalization or a lower temperature, such
that it lies outside of the PCA's bandpass. Furthermore, an excess is
visible in the fit residuals around $6$~keV (Figure~\ref{fig:spectrum}
bottom). This is similar to the {\it NuSTAR} spectra and was
interpreted as a redshifted and broadened Fe K$\alpha$ emission line
produced by photoionized reflection (Degenaar et al. 2017; Keek et
al. 2016). An additional Gaussian component can account for the
excess, but it does not significantly alter $\Gamma$.

\section{Discussion and Summary}

\label{sec:discussion} We have presented compelling evidence that 
\source{} harbors a 163.65 Hz pulsar. This is the lowest spin
frequency of the known AMXPs, where all others have $\nu\ge
182\ \mathrm{Hz}$ (e.g., Patruno \& Watts 2012).  While the present
data do not allow a precise measurement of the orbital period, there
is strong evidence for pulse frequency (and phase) evolution
consistent with circular orbital motion of the neutron star.  We can
find acceptable circular orbits with periods longward of $\approx 20$
minutes, however, periods shorter than this are disfavored, and we
determined a $90\%$ confidence lower limit on the orbital period of 17
minutes.

To place a limit on the magnetic field strength, we compare the
corotation radius, $R_\mathrm{co}$, and the magnetospheric radius,
$R_\mathrm{m}$ (e.g., Wijnands \& Van der Klis 1998). The former
refers to the radius where the gravitational infall is balanced by the
centrifugal force given the star's spin frequency:
$R_\mathrm{co}/R_\mathrm{g}=c^2
(GM\omega)^{-2/3}=27.0843(\frac{M}{1.4M_\odot})^{-2/3}$. The
magnetospheric radius is the location where the dipole magnetic
pressure equals the ram pressure of the infalling material:
$R_\mathrm{m}/R_\mathrm{g}=c^2(\frac{B^4R^{12}}{8G^8M^8\dot{M}^2})^{1/7}=
13.3(\frac{B}{10^8\mathrm{G}})^{4/7}(\frac{M}{1.4M_\odot})^{-8/7}
(\frac{R}{10\mathrm{km}})^{11/7}$, where we used a mass accretion rate
of $0.6\%\dot{M}_\mathrm{Edd}$ with $\dot{M}_\mathrm{Edd}=2\times
10^{-8}(\frac{R}{10\mathrm{km}})M_\odot \mathrm{yr}^{-1}$. For
accretion to occur, $R_\mathrm{m}$ must not exceed $R_\mathrm{co}$,
which places a constraint on the magnetic field strength: $B\le
3.5\times 10^8\,\mathrm{G}\,(\frac{M}{1.4M_\odot})^{5/6}(\frac{R}
{10\mathrm{km}})^{-11/4}$.  This is similar to the inferred magnetic
field strengths of other AMXPs, such as J1808 (Wijnands \& Van der
Klis 1998). For that source, $R_\mathrm{co}$ lies close to the neutron
star (see also Cackett et al. 2010), where relativistic corrections
are important (Psaltis \& Chakrabarty 1999). However, X-ray reflection
places the inner disk in \source{} at a distance of $R_\mathrm{in}\sim
10^2\ R_\mathrm{g}$ (Degenaar et al. 2017; Keek et al. 2016). Within
the large observational error on $R_\mathrm{in}$, it is consistent
with the above derived $R_\mathrm{co}$. Therefore, the relatively
strong magnetic field that presumably produces the pulsations in
\source{}, may also be responsible for truncating the accretion disk
at a substantial distance from the neutron star (Degenaar et
al. 2016).

If the magnetic field controls the flow of matter from $\sim
27\ R_\mathrm{g}$ down to the neutron star surface, this may have
important consequences for the accretion geometry in that region. The
reflection signal detected in the persistent emission by {\it NuSTAR}
in 2015 gives further hints about the geometry (Degenaar et al. 2016;
Keek et al. 2016). The inferred ionization parameter is large, even
though the illuminating accretion flux is low. This may indicate that
reflection occurs off material with a lower density than expected for
the inner disk. Furthermore, despite the reduction in area due to
truncation, the reflection fraction is substantial. Therefore,
reflection could occur off low-density gas near the truncation radius
where the magnetic field disrupts the disk (e.g, Ballantyne et
al. 2012). This material could subtend a larger angle and present a
substantial reflection surface.

A next step in understanding more clearly the accretion geometry would
be to accurately measure the orbital period. Several AMXPs are in
so-called ultra compact binaries (UCXBs), whereas others have longer,
few-hour periods. The orbital period can also be an indicator of the
composition of the accreted material, as hydrogen-rich donors may not
be able to physically reside within very compact systems. An example
is the UCXB 4U 1820-30, which must host a degenerate donor.  This is
important for the interpretation of the two thermonuclear X-ray bursts
(Keek et al. 2016; Degenaar et al. 2013) from \source{}. As we have
described, the {\it RXTE}/PCA observation was too short to accurately
determine the orbitial period; therefore, future timing observations
are needed, for example, with the {\it Neutron Star Interior
  Composition Explorer} (NICER; Gendreau et al. 2012) which is
scheduled for launch in 2017.

\label{sec:conclusions}

\acknowledgments

L.K. and T.S. acknowledge support by NASA under award number
NNG06EO90A. L.K. thanks the International Space Science Institute in
Bern, Switzerland and the National Science Foundation under grant
No. PHY-1430152 (JINA Center for the Evolution of the Elements) for
supporting events that benefited this work. We thank the anonymous
referee for a helpful review.

\newpage

%
\section*{References}

\begin{enumerate}

\item []{} Arnaud, K.~A. 1996, in Astronomical Society of the 
Pacific Conference Series, Vol. 101, Astronomical Data Analysis
Software and Systems V, ed. G.~H.  Jacoby \& J.~Barnes, 17

\item []{} Ballantyne, D.~R., Purvis, J.~D., Strausbaugh, R.~G., \& Hickox, R.~C. 2012, \apjl, 747, L35

\item []{} Bradt, H.~V. and {Rothschild}, R.~E. and {Swank}, J.~H. 1993, \aaps, 97, 355

\item []{} Buccheri, R., Bennett, K., Bignami, G.~F., et al.\ 1983, \aap, 128, 245 

\item[]{} Cackett, E.~M., {Miller}, J.~M., {Ballantyne}, D.~R., 
	{Barret}, D., {Bhattacharyya}, S., {Boutelier}, M., {Miller},
	M.~C., {Strohmayer}, T.~E., \& {Wijnands}, R., 2010, \apj,
	720, 205

\item[]{} {Chakrabarty}, D. and {Morgan}, E.~H. 1998, \nat, 394, 346

\item[]{} {Churazov}, E. and {Sunyaev}, R. and {Revnivtsev}, M. et al. 2007, \aap, 467, 529

\item[]{} {Degenaar}, N., {Miller}, J.~M., {Wijnands}, R., {Altamirano}, D., \& {Fabian},
  A.~C. 2013, \apjl, 767, L37

\item []{} Degenaar, N.,
  Pinto, C., Miller, J.~M., et al.\ 2017, \mnras, 464, 398

\item []{} {Gendreau}, K.~C., {Arzoumanian}, Z., \& {Okajima}, T. 2012, SPIE, 8443, 844313

\item []{} Keek, L., Iwakiri, W., Serino, M., et al.\ 2016, arXiv:1610.07608 

\item []{} Markwardt,
  C.~B., Swank, J.~H., Strohmayer, T.~E., in 't Zand, J.~J.~M., \&
  Marshall, F.~E.\ 2002, \apjl, 575, L21

\item []{} Negoro, H., Serino, M., Sasaki, R., et al.\ 2015, The Astronomer's Telegram, 8241

\item []{} Patruno, A., \& Watts, A.~L. 2012, arXiv:1206.2727

\item []{} Psaltis, D., \& Chakrabarty, D. 1999, ApJ, 521, 332

\item []{} {Remillard}, R.~A. and {Levine}, A.~M. 2008, ATel, 1853

\item []{} {Ricci}, C. and {Beckmann}, V. and {Carmona}, A. and {Weidenspointner}, G. 2008 , ATel, 1840

\item []{} Sanna, A., Papitto, A., Burderi, L., et al.\ 2016, arXiv:1611.02995 

\item []{} Strohmayer, T.~E., \& Markwardt, C.~B.\ 2002, ApJ, 577,
337

\item []{} van der Klis, M.\ 1989, NATO Advanced Science Institutes (ASI) Series C, 262, 27

\item []{} Wijnands, R., \& van der Klis, M.\ 1998, \nat, 394, 344 

\item []{} {Wilms} J.,  {Allen} A.,   {McCray} R.,  2000, \apj, 542, 914

\end{enumerate}

\newpage


\begin{figure*}
\begin{center}
\includegraphics[scale=0.75]{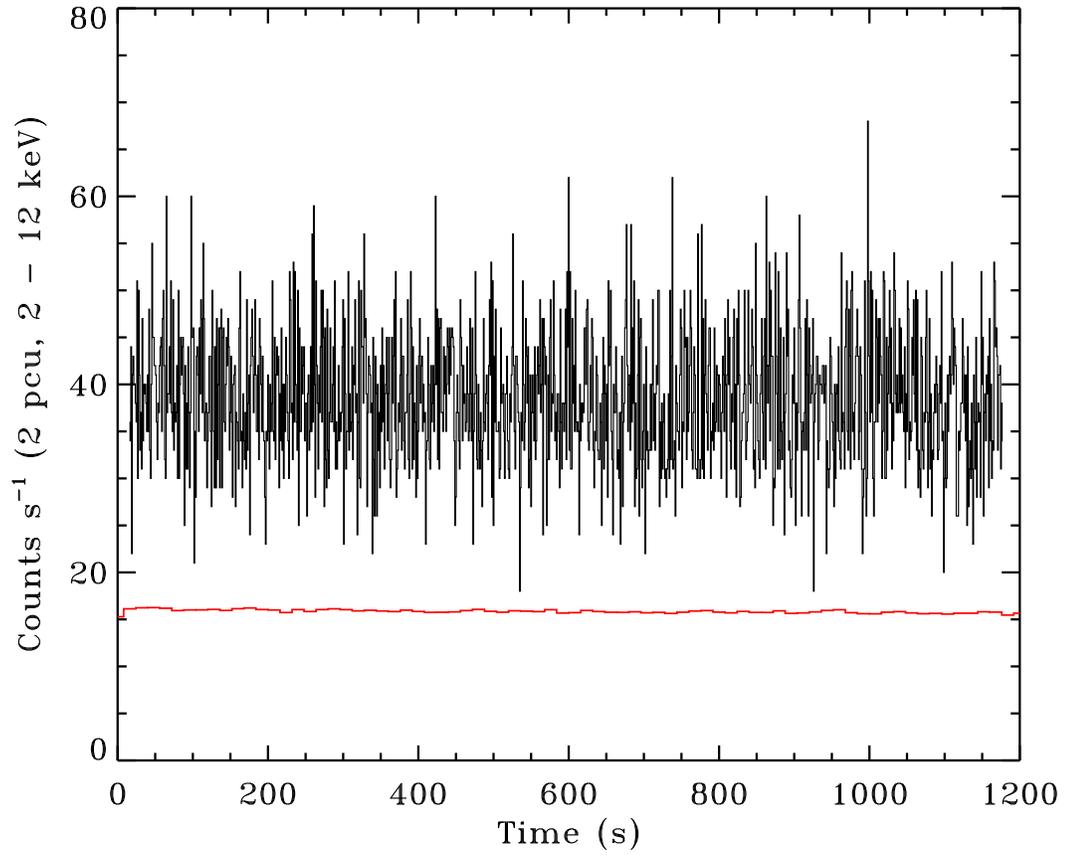}%
\end{center}
\caption{\label{fig:lc} Light curve of \source{} from {\it RXTE}
  PCA observations obtained on 2008 May 3. Data are the summed
  counting rates in 1 s bins in the 2 - 12 keV band from PCU 0 and 2
  (0-4 numbering scheme). The red histogram shows the background
  estimated from {\it pcabackest}. Time zero corresponds to
  12:58:39.866 UTC on the above date. }
\end{figure*}


\begin{figure*}
\begin{center}
\includegraphics[scale=0.75]{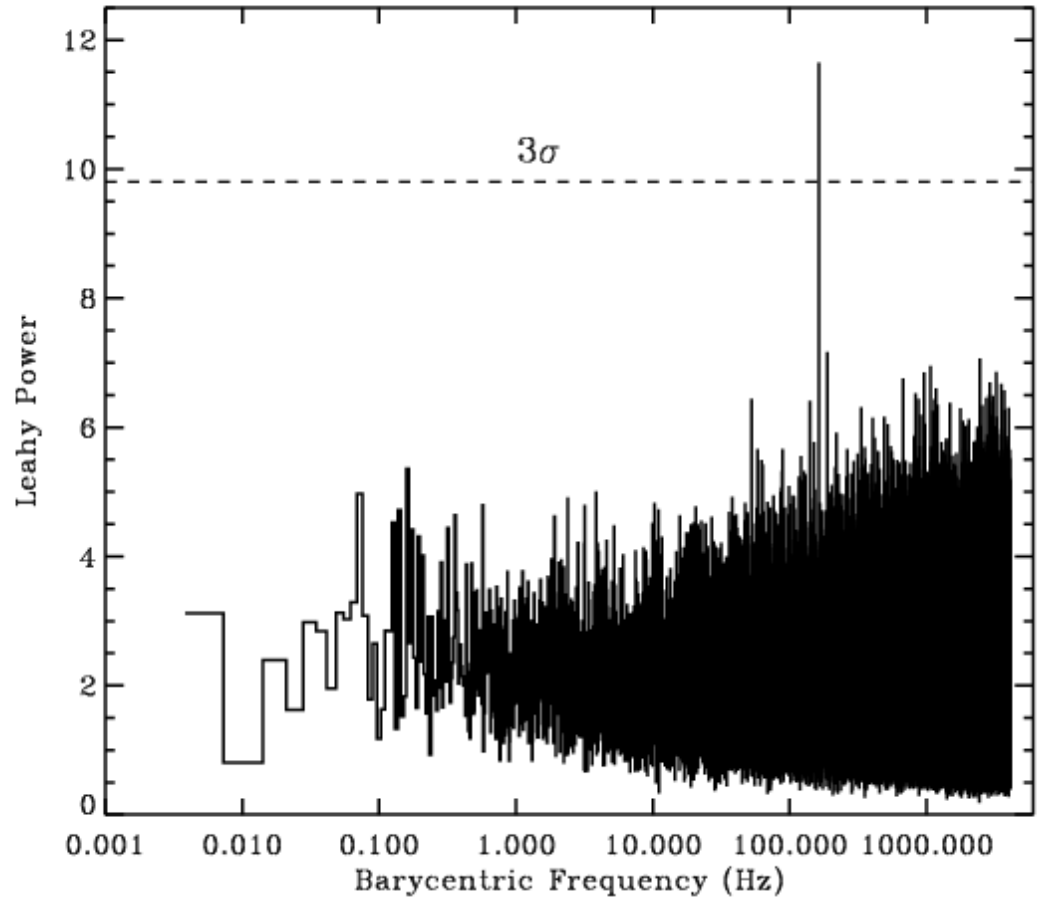}%
\end{center}
\caption{\label{fig:discover_pds} Power spectrum of \source{} as a
  function of frequency (Barycentric) computed from 2 to 12 keV events
  and rebinned by a factor of 8. The 163.65 Hz pulsar peak is evident
  just above 100 Hz.  The horizontal dashed line indicates the
  $3\sigma$ detection level. See \S 2 for a detailed discussion of the
  pulsation search. }

\end{figure*}


\begin{figure*}
\begin{center}
\includegraphics[scale=0.75]{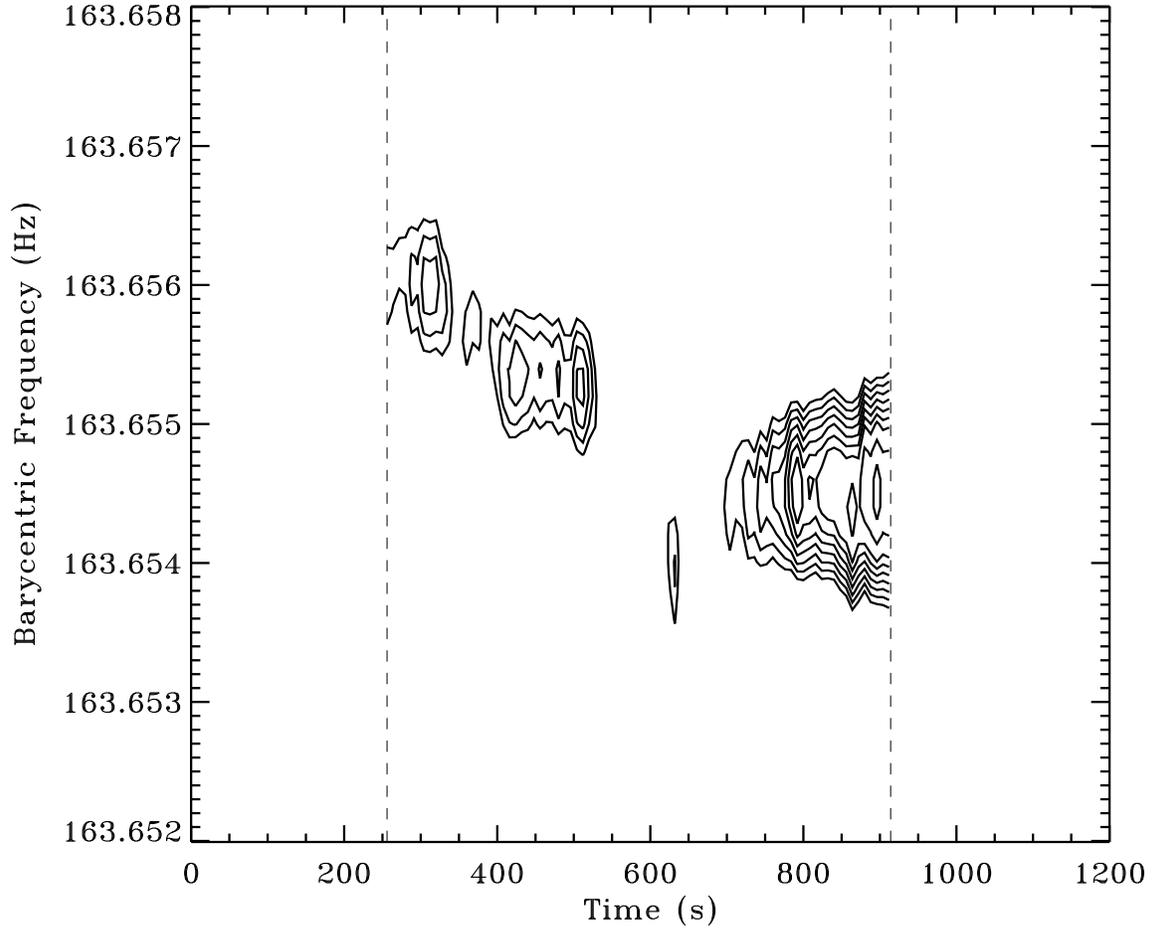}%
\end{center}
\caption{\label{fig:dynz} Dynamic power spectrum computed from 2 to 12
  keV events. Contours as a function of time and frequency are drawn
  at Leahy power values of 20, 22, 24, 26, 28, 30, 32, 36, and 40. We
  used 512 s intervals to compute power spectra, and computed a new
  spectrum by sliding the interval by 16 s. The contours are drawn at
  the times corresponding to the center of each interval used to
  compute a power spectrum. Dashed vertical lines appear at 256 s and
  $T_{obs}$-256 s, marking the centers of the first and last intervals
  used to compute spectra, respectively. Time zero corresponds to
  12:58:39.866 UTC on 2008 May 3. }
\end{figure*}


\begin{figure*}
\begin{center}
\includegraphics[scale=0.75]{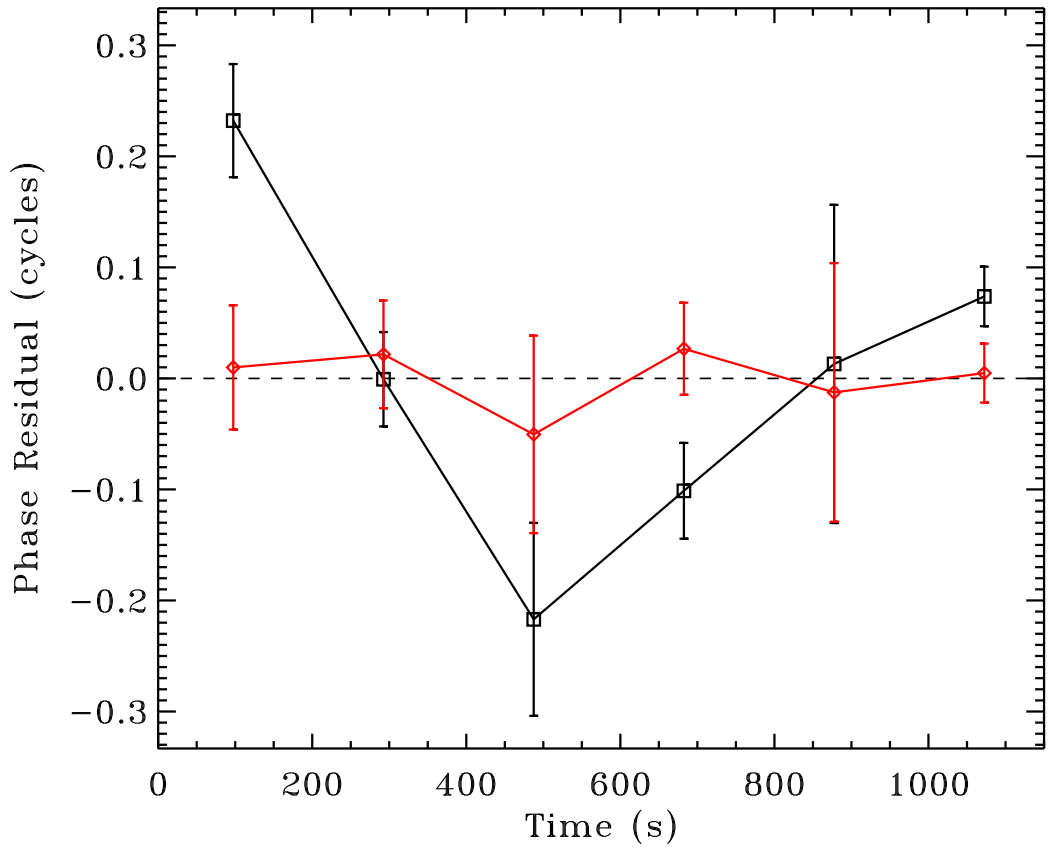}%
\end{center}
\caption{\label{fig:phase_resids} Results of coherent phase timing
  analysis. Phase residuals (cycles) vs. time from two different
  models for the frequency (and phase) evolution are shown. The black
  squares show the best-fitting, constant frequency model, which does
  not provide a good fit to the data, and shows a systematic trend
  indicative of spin-down of the pulsation frequency.  The red symbols
  show a statistically acceptable circular orbit model with a period
  of 30 minutes. Time zero corresponds to 12:58:39.866 UTC on 2008 May
  3. }
\end{figure*}

\begin{figure*}
\begin{center}
\includegraphics[scale=1.4]{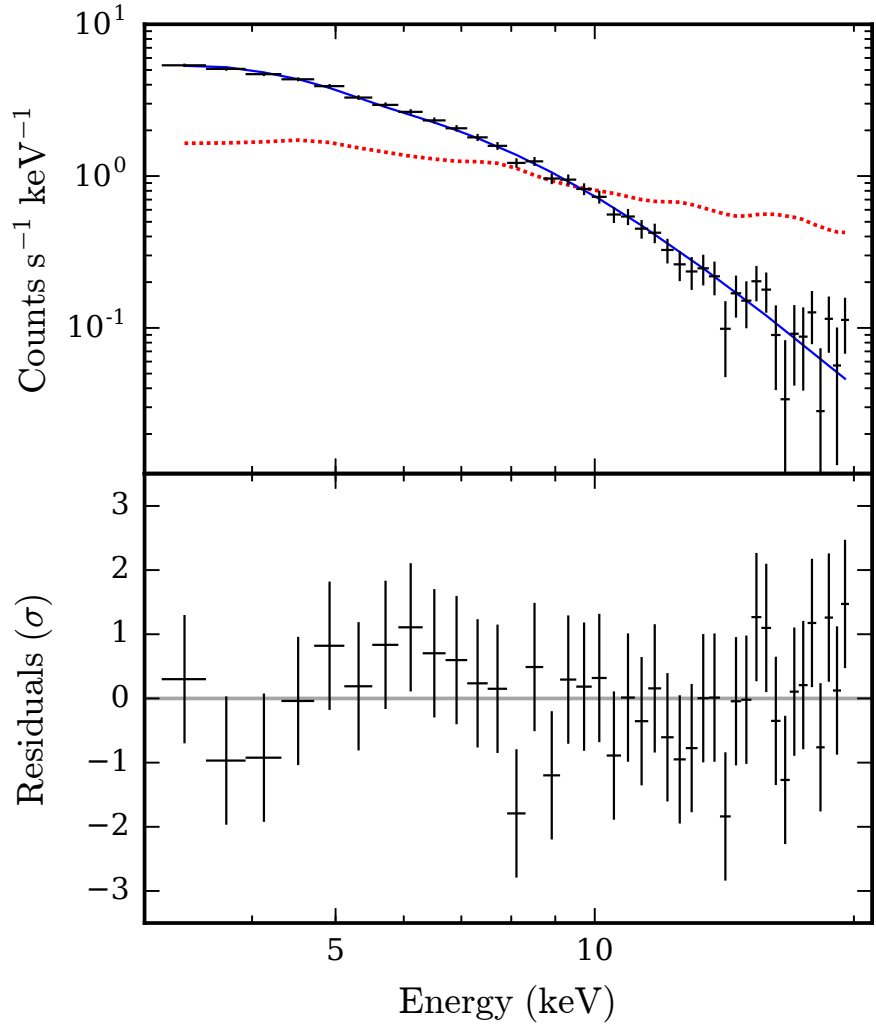}%
\end{center}
\caption{\label{fig:spectrum} PCA spectrum as a function of energy and
  the best-fitting absorbed power-law model (top; model in blue). The
  background (red curve) dominates the signal at high energies. A
  broad excess around $6$~keV is visible in the fit residuals
  (bottom), which may be due to photoionized reflection.}
\end{figure*}

\end{document}